\documentclass[conference]{IEEEtran}
\IEEEoverridecommandlockouts 

\usepackage{cite}
\usepackage{graphicx}
\usepackage{amsmath,amssymb,amsfonts}
\DeclareMathOperator*{\argmax}{arg\,max}
\usepackage{algorithm}
\usepackage{algorithmic}
\usepackage{textcomp}
\usepackage{xcolor}
\usepackage{hyperref} 
\usepackage{booktabs}
\usepackage{multirow}
\usepackage{subcaption}
\usepackage{listings}
\usepackage{xspace}
\usepackage{pifont} 
\usepackage{orcidlink}

\lstdefinelanguage{yaml}{
  keywords={true,false,null,y,n},
  keywordstyle=\color{blue}\bfseries,
  basicstyle=\ttfamily\footnotesize,
  identifierstyle=\color{black},
  sensitive=false,
  comment=[l]{\#},
  commentstyle=\color{gray}\itshape,
  stringstyle=\color{teal},
  morestring=[b]',
  morestring=[b]"
}

\newcommand{\hubscan}{\textsc{Adversarial Hubness Detector}\xspace}
\newcommand{\etal}{\textit{et al.}\xspace}

\hypersetup{
  colorlinks=true,
  linkcolor=blue,
  citecolor=blue,
  urlcolor=blue,
  pdftitle={Adversarial Hubness Detector: Detecting Hubness Poisoning in Retrieval-Augmented Generation Systems},
  pdfauthor={Manish Bhatt, Ronald F. Del Rosario, Vineeth Sai Narajala, Idan Habler}
}

\begin{document}

\title{\Large \bf Adversarial Hubness Detector: Detecting Hubness Poisoning in Retrieval-Augmented Generation Systems}

\author{
\IEEEauthorblockN{Idan Habler\,\orcidlink{0000-0003-3423-5927}}
\IEEEauthorblockA{\textit{Cisco | OWASP} \\
ihabler@cisco.com}
\and
\IEEEauthorblockN{Vineeth Sai Narajala\,\orcidlink{0009-0007-4553-9930}}
\IEEEauthorblockA{\textit{Cisco | OWASP} \\
vineeth.sai@owasp.org}
\and
\IEEEauthorblockN{Stav Koren\,\orcidlink{0009-0007-7322-9083}}
\IEEEauthorblockA{\textit{Tel Aviv University } \\
stavk@mail.tau.ac.il}
\and
\IEEEauthorblockN{Amy Chang\,\orcidlink{0009-0005-9910-1673}}
\IEEEauthorblockA{\textit{Cisco} \\
changamy@cisco.com}
\and
\IEEEauthorblockN{Tiffany Saade\,\orcidlink{0009-0009-2851-7144}}
\IEEEauthorblockA{\textit{Cisco} \\
tsaade@cisco.com}
}

\maketitle
\thispagestyle{plain}
\pagestyle{plain}

\begin{abstract}
Retrieval-Augmented Generation (RAG) systems are essential to contemporary AI applications, allowing large language models to obtain external knowledge via vector similarity search. Nevertheless, these systems encounter a significant security flaw: \textit{hubness} - items that frequently appear in the top-$k$ retrieval results for a disproportionately high number of varied queries. These hubs can be exploited to introduce harmful content, alter search rankings, bypass content filtering, and decrease system performance.

We introduce \hubscan, an open-source security scanner that evaluates vector indices and embeddings to identify hubs in RAG systems. \hubscan presents a multi-detector architecture that integrates: (1) robust statistical hubness detection utilizing median/Median Absolute Deviation (MAD)-based z-scores, (2) cluster spread analysis to assess cross-cluster retrieval patterns, (3) stability testing under query perturbations, and (4) domain-aware and modality-aware detection for category-specific and cross-modal attacks. Our solution accommodates several vector databases (FAISS, Pinecone, Qdrant, Weaviate) and offers versatile retrieval techniques, including vector similarity, hybrid search, and lexical matching with reranking capabilities.

We evaluate \hubscan on Food-101, MS-COCO, and FiQA adversarial hubness benchmarks constructed using state-of-the-art gradient-optimized and centroid-based hub generation methods. \hubscan achieves 90\% recall at a 0.2\% alert budget and 100\% recall at 0.4\%, with adversarial hubs ranking above the 99.8th percentile. In testing, domain-scoped scanning recovered 100\% of targeted attacks that evaded global detection. Production validation on 1M real web documents from MS MARCO demonstrates significant score separation between clean documents and adversarial content.
Our work provides a practical, extensible framework for detecting hubness threats in production RAG systems.
Our open-source implementation is available at \url{https://github.com/cisco-ai-defense/adversarial-hubness-detector}.
\end{abstract}

\section{Introduction}
\label{sec:introduction}
Retrieval-Augmented Generation (RAG) systems have become an essential framework for grounding large language models (LLMs) in external knowledge \cite{lewis2020retrieval}. RAG systems can deliver precise, current responses by extracting relevant items from a vector database prior to generation, eliminating the need for continuous model retraining. This framework facilitates several applications, including corporate knowledge repositories, customer support systems, and AI assistants across multiple sectors.

The fundamental mechanism of RAG systems is based on vector similarity search within high-dimensional embedding spaces. Upon submission of a query, it is transformed into a vector representation, and the system obtains the $k$ most common items from a vector index. The collected texts are subsequently supplied as context to an LLM for answer generation. This architecture presents considerable benefits, although it also creates a crucial attack surface: the vector embedding space.

\subsection{The Hubness Threat}

\textit{Hubness} is a recognized occurrence in high-dimensional spaces where specific locations frequently serve as nearest neighbors to several other points \cite{radovanovic2010hubs}. In information retrieval situations, this results in certain items appearing in the top-$k$ results more frequently than anticipated. Hubness can arise organically; nevertheless, it also implies a significant security weakness that may be intentionally exploited.

Zhang \etal~\cite{zhang2024adversarial} demonstrated that attackers can craft embeddings that become ``hubs'' appearing in retrieval results for thousands of semantically unrelated queries. In their experiments on text-caption-to-image retrieval, a single hub generated using only 100 random queries was retrieved as the top-1 result for over 21,000 out of 25,000 test queries---far exceeding expected behavior for normal items.

Such \emph{hub injection} attacks enable a form of universal \textbf{retrieval poisoning}: by inserting a malicious item into a database, an adversary can force irrelevant or harmful content to appear in a wide array of search results \cite{zhang2024adversarial}.

\subsection{Real-World Attack Incidents}

Real-world incidents underscore the feasibility and severity of retrieval poisoning:

\begin{itemize}
    \item \textbf{Microsoft 365 Copilot Poisoning}: Zenity Labs demonstrated that ``all you need is one document'' to reliably mislead a production Copilot system \cite{zenity2025zeroclick}. A single crafted document made the Copilot confidently answer queries with attacker-chosen false facts.
    
    \item \textbf{GeminiJack Attack}: Noma Security's exploit showed that a single shared Google Doc with embedded instructions caused Google's Gemini AI search to exfiltrate months of private emails and documents~\cite{levi2025geminijack}, demonstrating a zero-click data breach through retrieval poisoning.
    
    \item \textbf{Content Injection}: Hubs enable attackers to inject malicious, misleading, or spam content that appears in responses to diverse queries.
    
    \item \textbf{Indirect Prompt Injection}: By placing prompt injection payloads in hub items, attackers can manipulate LLM behavior across many user queries \cite{greshake2023indirect}.
    
    \item \textbf{The Promptware Kill Chain}: Recent research by \cite{nassi2026promptware} formalizes these exploits as part of a multi-step malware lifecycle. They demonstrate how retrieval poisoning serves as the primary \textit{delivery vector} for \textbf{Promptware}—AI-native malware that utilizes RAG to move from initial access to lateral movement and data exfiltration and gain persistency.
\end{itemize}
These examples underscore a critical weakness in RAG architectures: the absence of a resilient poison detection layer to capture harmful payloads prior to their integration into the model's environment. In the absence of such identification, the RAG system autonomously compromises itself by considering contaminated external inputs as reliable information.

\subsection{Challenges in Detection}

Overcoming hubness is difficult as the concept of the attack exploits intrinsic characteristics of embedding spaces. Previous methods for decreasing \emph{natural} hubness (e.g., similarity normalization \cite{conneau2018word, wang2023balance}) solely diminish hubs that are globally frequent neighbors. An adaptive adversary can establish a \textit{domain-specific hub} that activates solely for inquiries on a single subject, circumventing overarching defenses \cite{zhang2024adversarial}.

Key detection challenges include:

\begin{enumerate}
    \item \textbf{Statistical Robustness}: Hubs demonstrate hub rates exceeding the median by 5-10+ standard deviations, necessitating the use of robust statistical procedures that are resilient to the impact of outliers.
    
    \item \textbf{Domain-Specific Attacks}: Advanced attackers establish hubs that focus on particular semantic categories (e.g., "medical advice" or "financial reports"), circumventing worldwide detection while exerting significant influence inside their designated domain.
    
    \item \textbf{Cross-Modal Attacks}: In multimodal systems, items can be designed to seem related to queries in an alternate modality, taking advantage of modality boundaries that single-modality detection overlooks.
    
    \item \textbf{Semantic Mismatch}: Hubs get high similarity scores at the expense of genuine semantic alignment even though their content frequently diverges from the queries that yield them.
    
    \item \textbf{Retrieval Method Awareness}: Detection must be effective across several retrieval methods, including vector search, hybrid search, and reranking pipelines.
\end{enumerate}

\subsection{Contributions}

We present \hubscan, the first comprehensive detection system for hubness in RAG systems. Our contributions include:

\begin{enumerate}
    \item \textbf{Multi-Detector Architecture}: A pluggable detection framework combining hubness frequency analysis, cluster spread detection, stability testing, and deduplication---each targeting different aspects of hub behavior.
    
    \item \textbf{Robust Statistical Methods}: Novel application of median/MAD-based z-scores that maintains numerical stability for detecting statistical anomalies in hub rate distributions.
    
    \item \textbf{Domain-Aware Detection}: Methods for identifying hubs that focus on particular semantic categories via bucket analysis of specific RAG indices.
    
    \item \textbf{Modality-Aware Detection}: Cross-modal anomaly detection for multimodal systems, incorporating parallel retrieval and late fusion architectures.
    
    \item \textbf{Flexible Retrieval Support}: Integration with multiple vector databases and retrieval methods, including hybrid search and custom reranking algorithms through a plugin system and customizable interfaces.
    
    \item \textbf{Open-Source Implementation}: A production-ready scanner with adapters to the most popular RAG frameworks.
\end{enumerate}

Against SOTA hubs~\cite{zhang2024adversarial} (CLIP/Qwen3), \hubscan achieved \textbf{90\% recall} at a 0.2\% alert budget and \textbf{100\% recall} at 0.4\%. The Cluster Spread Detector provides critical lift (+10--20pp recall) for universal attacks. Production validation on 1 million MS MARCO~\cite{nguyen2016ms} passages demonstrated \textbf{5.8$\times$ score separation} between adversarial hubs and the 99\textsuperscript{th} percentile of clean data, with 0.1\% operational overhead, confirming web-scale deployment feasibility.

The rest of this paper is organized as follows: Section~\ref{sec:related_work} reviews related work. Section~\ref{sec:methodology} presents our detection methodology and background. Section~\ref{sec:detection} details the detection algorithms. Section~\ref{sec:domain_modality} describes domain-aware and modality-aware detection. Section~\ref{sec:retrieval} covers retrieval integration. Sections~\ref{sec:evaluation},\ref{sec:production_audit} presents our evaluation. Section~\ref{sec:conclusion} concludes.


\section{Related Work}
\label{sec:related_work}

\subsection{Hubness in High-Dimensional Spaces}

The hubness phenomenon was initially defined by Radovanović \etal~\cite{radovanovic2010hubs}, who illustrated that in high-dimensional spaces, specific points-termed \textbf{"hubs"}, frequently serve as nearest neighbors to numerous other points with disproportionate prevalence. Subsequent research investigated mitigation methods: Mutual Proximity \cite{schnitzer2012local} modifies distances according to local neighborhoods, while CSLS \cite{conneau2018word} standardizes similarities by employing average neighbor distances. Lin \etal~\cite{lin2025neighborretr} presented NeighborRetr to equilibrate hub centrality during training, and Wang \etal~\cite{wang2023balance} introduced Dual Bank Normalization (DBNorm) as a post-processing method. However, these methodologies concentrate on \textit{natural} hubness resulting from data distributions, rather than on adversarially constructed hubs.

\subsection{Hubness Attacks}

Zhang \etal~\cite{zhang2024adversarial} conducted the first extensive examination of hubness exploitation in multi-modal retrieval, demonstrating that adversaries can leverage the hubness phenomena to generate embeddings obtained from thousands of unrelated queries. Their research showed that a singular hub might prevail in top-1 outcomes for 84\% of test queries in text-to-image retrieval, with hubs generalizing far beyond their training queries (100 queries used to create a hub affecting 21,000+ queries). Existing hubness mitigation techniques provide limited defense against targeted hubs, which can be crafted to target specific concepts while evading global detection. Our research expands upon these findings to establish a \textbf{detection framework} explicitly aimed at hubs.

\textbf{Real-World Incidents}: Security research firms have documented production RAG vulnerabilities. Zenity Labs~\cite{zenity2025zeroclick} reported that ``all you need is one document'' to reliably mislead Microsoft 365 Copilot. Noma Security (GeminiJack) uncovered vulnerabilities where an indexed document with hidden instructions induced Google Gemini to leak confidential data---a zero-click data exfiltration attack. Greshake \etal~\cite{greshake2023indirect} demonstrated indirect prompt injection through the retrieval pathway. These examples highlight the potential harm when retrieval is impaired; our research focuses on attacks that function at the embedding level, utilizing continuous vector spaces.

\subsection{Common Retrieval Methods}

Wu \etal~\cite{wu2022retrievalguard} introduced RetrievalGuard, a demonstrably resilient 1-nearest neighbor image retrieval technique that preserves accuracy in the presence of adversarial perturbations. Hybrid search, which integrates vector similarity with lexical matching, has grown popular for enhancing retrieval quality \cite{bansal2023optimizing}. Reranking methodologies employing cross-encoders \cite{nogueira2019passage} offer secondary filtering. Our work integrates with these approaches to provide defense-in-depth.


\section{Methodology}
\label{sec:methodology}

\subsection{Threat Model and Problem Definition}
\label{subsec:methodology-threat-model}
We assume an adversary can introduce or modify an item in the retrieval system's index by applying perturbations to a base item (image, audio, document, etc.). The adversary's goal is to maximize the number of query embeddings that rank this item among the top results. They may target either \emph{universal hubness} (dominate retrieval for virtually all queries) or \emph{domain-specific hubness} (dominate queries related to a particular topic). The attacker does \textit{not} necessarily control query inputs; instead, they exploit embedding space geometry so that many legitimate user queries will naturally retrieve the adversarial item.

The attacker has the following capabilities: (1)~\textbf{Write Access} to insert or modify documents in the vector database; (2)~\textbf{Query Sampling} to sample queries from the target distribution; and (3)~\textbf{Embedding Knowledge} of the embedding model used by the target system. The attacker's goal is to inject hub documents that appear in top-$k$ results for a large fraction of user queries, carry malicious payloads, and evade detection by appearing statistically similar to normal content.

In our system, every gallery item is converted into a point in an embedding space, and searches work by retrieving the items located closest to the query. We define a "hub" as a gallery item that exploits this geometry to appear in the top results for a disproportionately large number of searches. \hubscan's objective is to scan the gallery to identify these crafted items and then either remove them or suppress their ranking to prevent them from dominating the results.

\subsection{Hubness Background}
\label{subsec:hubness-background}
In vector search, we assess a gallery item's impact by quantifying its frequency of appearance in the top results across an extensive array of queries. In high-dimensional data, the distribution is inherently skewed \cite{radovanovic2010hubs}: certain data points, referred to as "hubs," function as attractors and are retrieved significantly more often than the mean, whereas others, termed "anti-hubs," are seldom accessed.

A \textit{hub} is an entity that consistently ranks inside the top-$k$ results for a significant proportion of varied searches. Hubness inherently occurs in high-dimensional embedding spaces, but it can also be deliberately employed to construct \emph{adversarial hubs} that dominate in retrieval. As demonstrated by Zhang \etal~\cite{zhang2024adversarial}, adversarial hubs can be produced by gradient-based optimization: the algorithm selects a limited number of queries, refines an embedding that corresponds with their common semantic direction, and implements a constrained modification to a gallery item to ensure its retrieval for many queries.

Hubs exhibit several distinguishing characteristics: (1)~\textbf{Extreme Hub Rates} of 20--50\%+ compared to expected rates of 2--5\% for normal documents; (2)~\textbf{Cross-Cluster Spread}---retrieved by queries from many diverse semantic clusters; (3)~\textbf{Perturbation Stability}---maintaining high retrieval rates even under query perturbations; and (4)~\textbf{Statistical Anomaly}---hub z-scores typically exceeding 5--10 standard deviations above the median.

\subsection{System Architecture}
\label{subsec:methodology-architecture}
\begin{figure}[!t]
\centering
\includegraphics[width=1\linewidth]{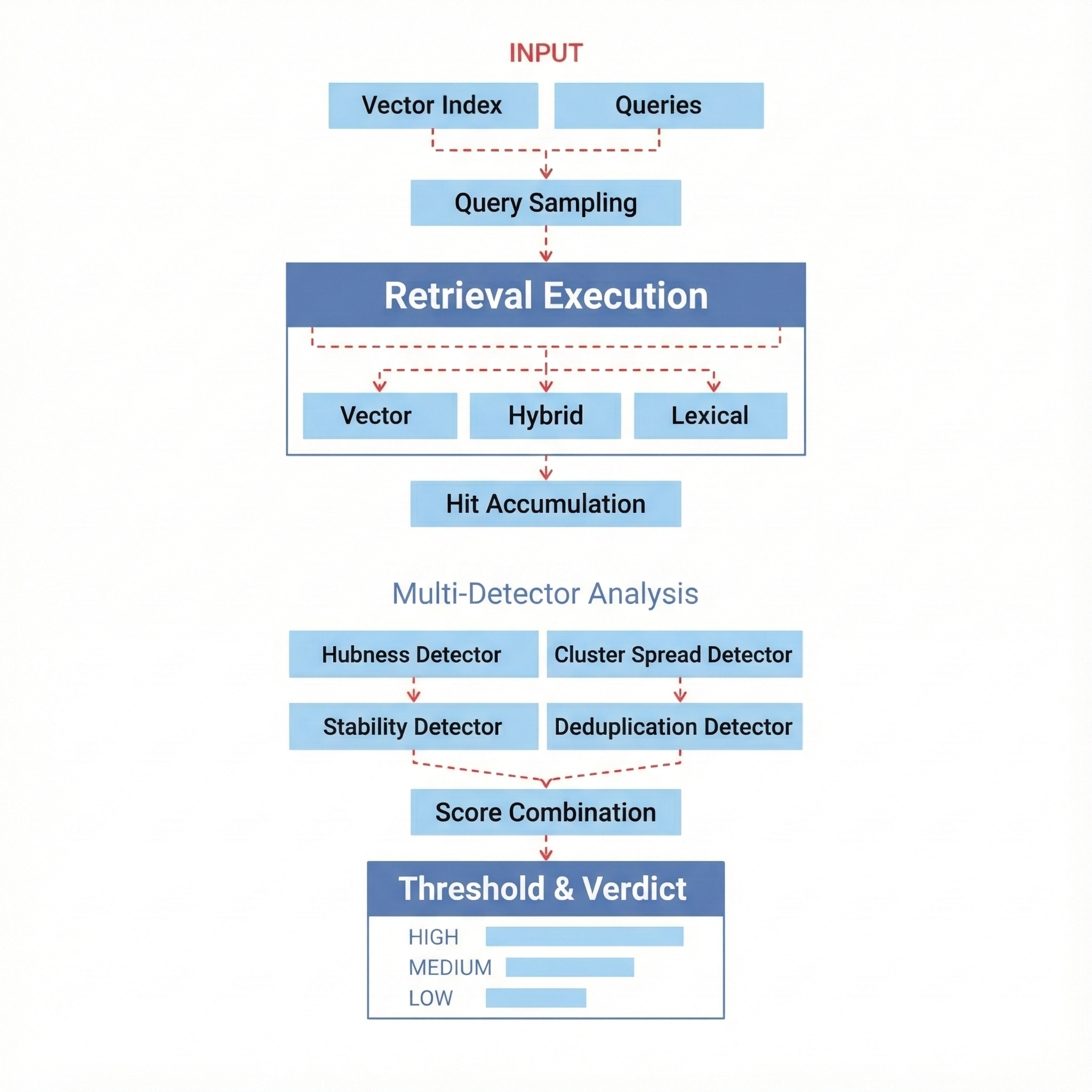}
\caption{Adversarial Hubness Detector detection pipeline overview showing the multi-stage process from input to verdict assignment.}
\label{fig:architecture}
\end{figure}

\hubscan implements a multi-detector architecture that analyzes vector indices through several complementary lenses. Figure~\ref{fig:architecture} illustrates the overall pipeline.

The detection process operates as follows:
\begin{enumerate}
    \item \textbf{Data Loading}: Load document embeddings, metadata, and vector index from supported backends (FAISS, Pinecone, Qdrant, Weaviate).
    \item \textbf{Query Sampling}: Generate or sample representative queries from the document distribution.
    \item \textbf{Retrieval Execution}: Execute $k$-NN queries using configurable retrieval methods, accumulating hit counts per document.
    \item \textbf{Detection}: Run multiple detectors analyzing different aspects of adversarial behavior.
    \item \textbf{Score Fusion}: Combine detector outputs using weighted scoring.
    \item \textbf{Verdict Assignment}: Classify documents as HIGH, MEDIUM, or LOW risk based on configurable thresholds.
\end{enumerate}

\subsection{Query Sampling Strategy}
\label{subsec:methodology-query-sampling}
Effective hubness detection requires representative queries that cover the semantic space of the document corpus. The query sampling employs a \textit{mixed sampling strategy} combining multiple approaches:
\begin{enumerate}
    \item \textbf{Cluster Centroids}: We apply MiniBatch K-Means clustering to the item embeddings and use cluster centroids as queries, ensuring queries are distributed across semantic regions.
    \item \textbf{Random Items Sampling}: We randomly sample item embeddings to serve as queries, ensuring queries reflect the actual item distribution.
    \item \textbf{Real Queries} (when available): For datasets with pre-existing query sets, we incorporate actual user queries to better reflect production query patterns.
\end{enumerate}

This mixed strategy integrates semantic coverage (via centroids) with distributional authenticity (by random sampling), providing detection across varied query patterns while preserving computing efficiency with MiniBatch K-Means.

\subsection{Robust Statistical Framework}
A key problem in hubness detection is that hubs function as extreme outliers, distorting conventional mean and variance computations. To address this, we standardize hub rates utilizing the MAD. This rigorous method produces consistent z-scores in the existence of formidable adversary samples, enabling us to accurately identify papers with scores surpassing 5 as potential threats.

\subsection{Weighted Hit Accumulation}
\label{subsec:methodology-weighted-hits}
A basic hub score quantifies the frequency with which a page appears in the top-$k$ retrieval results across various queries. This strategy, while successful as a baseline, implicitly presumes that all appearances contribute equally to hubness, which is not accurate in actuality. Documents that consistently occupy high-ranking positions or are located at minimal embedding distances demonstrate far more pronounced hub behavior than those that appear infrequently near the threshold.

We developed a \emph{weighted hit accumulation} system to measure this better. Instead of counting every match equally, we give more points to matches that appear at the top of the list or are very similar. This shows that powerful items dominate because they take the best spots, not just because they have been retrieved often.

\subsection{Multi-Detector Scoring}
\label{subsec:methodology-scoring}
Each detector produces a per-item suspicion score $s_i^{(d)}$, where higher values indicate more anomalous behavior. Since detectors may emit scores on different scales, we optionally apply a per-detector normalization $g_d(\cdot)$ to obtain comparable normalized scores $\tilde{s}_i^{(d)} \in [0,1]$. We then fuse signals via a weighted sum:
\begin{equation}
s_i^{\text{combined}} = \sum_{d} \alpha_d \, \tilde{s}_i^{(d)} \,,
\qquad \tilde{s}_i^{(d)} = g_d\!\left(s_i^{(d)}\right)
\end{equation}
where $\alpha_d$ are configurable weights controlling each detector's contribution. Default weights are: Hubness z-score (1.0), Domain Hub (0.5), Cross-Modal Penalty (0.5), Cluster Spread (0.3), and Stability (0.2).

\subsection{Verdict Classification}
\label{subsec:methodology-verdicts}
Items are classified into risk levels based on combined scores:
\begin{itemize}
    \item \textbf{HIGH}: $h \geq 99^{\text{th}}$ percentile
    \item \textbf{MEDIUM}: $h \geq 98^{\text{th}}$ percentile
    \item \textbf{LOW}: Below MEDIUM thresholds
\end{itemize}
These thresholds are tunable and can be adjusted based on operational requirements (higher precision vs. higher recall).

\subsection{Mitigation via Re-ranking}
Once a candidate hub is confirmed, the defender can mitigate its impact through removal, quarantine, or a \textit{re-ranking filter} that pushes flagged items down in results. One approach subtracts a penalty from the similarity score of flagged items. Although hubness reduction transforms alone are insufficient against adaptive, domain-specific hubs \cite{zhang2024adversarial}, in combination with explicit detection they can dramatically reduce a hub's visibility with minimal impact on normal results.


\section{Detection Algorithms}
\label{sec:detection}

\begin{figure}[!t]
\centering
\includegraphics[width=1\linewidth]{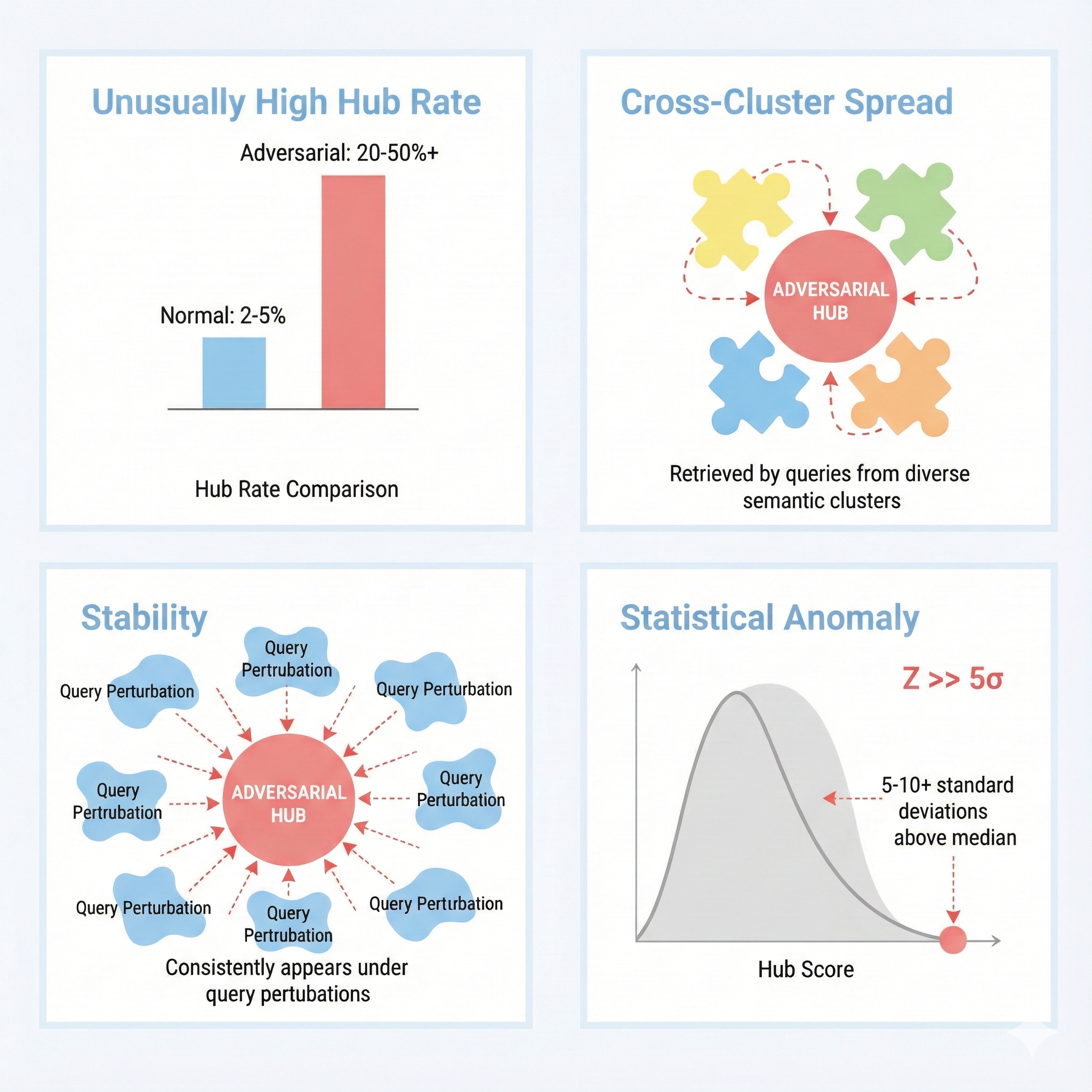}
\caption{Key detection metrics and their interpretation: Hub z-score measures statistical anomaly, cluster entropy captures cross-cluster spread, stability indicates robustness to perturbations, and combined scores provide holistic risk assessment.}
\label{fig:detection_metrics}
\end{figure}

\hubscan implements four complementary detectors, each targeting different aspects of hub behavior. This section details the algorithms and their underlying rationale, Figure~\ref{fig:detection_metrics} illustrates key detection metrics.

\subsection{Hubness Detector}

The core hubness detector implements reverse $k$-NN frequency analysis with bucketed accumulation for efficient processing.

\begin{algorithm}
\caption{Hubness Detection}
\begin{algorithmic}[1]
\REQUIRE Index $\mathcal{I}$, queries $Q$, documents $D$, $k$
\STATE Initialize accumulator $A$ with $|D|$ documents
\FOR{each query $q \in Q$}
    \STATE $\text{neighbors}, \text{distances} \leftarrow \text{kNN}(\mathcal{I}, q, k)$
    \FOR{$r = 1$ to $k$}
        \STATE $d_{\text{idx}} \leftarrow \text{neighbors}[r]$
        \STATE $w \leftarrow w_{\text{rank}}(r) \cdot w_{\text{dist}}(\text{distances}[r])$
        \STATE $A.\text{add\_hit}(d_{\text{idx}}, w)$
    \ENDFOR
\ENDFOR
\STATE $h \leftarrow A.\text{compute\_hub\_rates}()$
\STATE $z, \mu, \sigma \leftarrow \text{robust\_zscore}(h)$
\RETURN $z$
\end{algorithmic}
\end{algorithm}

The bucketed accumulator tracks hits per document with optional bucketing by concept or modality, enabling both global and fine-grained analysis from a single pass over queries. The primary output is the robust z-score vector indicating how many ``standard deviations'' each document's hub rate deviates from the median. Gallery items in the $99^{\text{th}}$ percentile are highly anomalous---for comparison, in a normal distribution, $z = 6$ corresponds to approximately 1 in 500 million probability.

\subsection{Cluster Spread Detector}

Hubs are designed to capture queries from multiple semantic regions. The cluster spread detector measures this cross-cluster retrieval pattern.

\begin{algorithm}
\caption{Cluster Spread Detection}
\begin{algorithmic}[1]
\REQUIRE Queries $Q$, query embeddings $E_Q$, $k$, $n_{\text{clusters}}$
\STATE Cluster queries: $C \leftarrow \text{KMeans}(E_Q, n_{\text{clusters}})$
\FOR{each query $q_j \in Q$}
    \STATE $c_j \leftarrow C.\text{predict}(q_j)$
    \STATE $\text{neighbors} \leftarrow \text{kNN}(\mathcal{I}, q_j, k)$
    \FOR{each $d_{\text{idx}} \in \text{neighbors}$}
        \STATE $\text{cluster\_hits}[d_{\text{idx}}][c_j] \mathrel{+}= 1$
    \ENDFOR
\ENDFOR
\FOR{each document $d_i$}
    \STATE $p \leftarrow \text{normalize}(\text{cluster\_hits}[d_i])$
    \STATE $s_i \leftarrow \text{entropy}(p) / \log(n_{\text{clusters}})$
\ENDFOR
\RETURN $s$ (normalized entropy scores)
\end{algorithmic}
\end{algorithm}

We categorize search queries into semantic clusters utilizing MiniBatch K-Means and assess the extent of a gallery item's distribution among them. By computing a normalized Shannon entropy~\cite{shannon1948mathematical} score reflecting the diversity of hits across query clusters, we differentiate between authentic content and adversarial hubs. A high normalized entropy score (close to 1.0) signifies that the gallery item is distributed evenly across numerous unrelated semantic clusters---indicative of a hub intended for extensive coverage. Normal gallery items exhibit lower entropy, focusing their hits within their specific subject area.

\subsection{Stability Detector}

Hubs maintain high retrieval rates under query perturbations due to their central positioning in embedding space. The stability detector exploits this characteristic.

\begin{algorithm}
\caption{Stability Detection}
\begin{algorithmic}[1]
\REQUIRE Top candidate documents $D_{\text{top}}$, original queries $Q$
\STATE Initialize stability scores $s$
\FOR{each candidate $d_i \in D_{\text{top}}$}
    \STATE $\text{original\_hits} \leftarrow \text{count\_hits}(d_i, Q)$
    \STATE $\text{perturbed\_hits} \leftarrow []$
    \FOR{$p = 1$ to $n_{\text{perturbations}}$}
        \STATE $Q' \leftarrow Q + \mathcal{N}(0, \sigma^2 I)$ \COMMENT{Add Gaussian noise}
        \STATE $Q' \leftarrow \text{normalize}(Q')$
        \STATE $\text{perturbed\_hits}.\text{append}(\text{count\_hits}(d_i, Q'))$
    \ENDFOR
    \STATE $s_i \leftarrow \text{mean}(\text{perturbed\_hits}) / \text{original\_hits}$
\ENDFOR
\RETURN $s$
\end{algorithmic}
\end{algorithm}

We apply Gaussian perturbations with $\sigma = 0.01$ (configurable) to query embeddings. Gaussian noise provides \textit{isotropic perturbations} that uniformly explore the local embedding neighborhood. Gallery items maintaining high hit rates across perturbations receive higher stability scores. High stability ($s \approx 1$) suggests the document is geometrically central, a hallmark of hubs. Normal gallery items show lower stability as perturbed queries drift to other topics. For efficiency, we only test the top-$X$ candidates (default: top 200) identified by the hubness detector.

\subsection{Deduplication Detector}

Attackers may inject multiple near-duplicate hubs to increase coverage or evade single-document thresholds. The deduplication detector identifies such clusters using: (1)~\textbf{Exact Text-Hash Grouping} when a \texttt{text\_hash} field is available; (2)~\textbf{Embedding-Based Near-Duplicate Detection} via nearest-neighbor search with a distance threshold; and (3)~\textbf{Boilerplate Suppression} to downweight large duplicate clusters (often templated content). Gallery items in duplicate clusters receive adjusted scores based on cluster size.

\subsection{Detector Compatibility}

Not all detectors are applicable to all retrieval methods. The Hubness and Deduplication detectors work with vector, hybrid, and lexical retrieval. The Cluster Spread and Stability detectors require semantic query embeddings, making them incompatible with pure lexical search. \hubscan automatically skips incompatible detectors based on the configured retrieval method.

\subsection{Score Interpretation}

The combined scoring produces interpretable risk assessments: \textbf{High Hub Z-Score} identifies extremely anomalous documents appearing in significantly more queries than statistically expected. \textbf{High Cluster Entropy} reflects wide cross-cluster spread across diverse and unrelated topics. \textbf{High Stability} indicates robustness to perturbations, where documents maintain retrieval dominance under query variations. The \textbf{Combined Score} provides a weighted aggregation for holistic risk assessment.

Default thresholds are calibrated for high precision (minimizing false positives) while maintaining strong recall on known hub patterns.


\section{Domain-Aware and Modality-Aware Detection}
\label{sec:domain_modality}

\begin{figure}[h!]
\centering
\includegraphics[width=1\linewidth, trim={0cm 8cm 0cm 0cm}, clip]{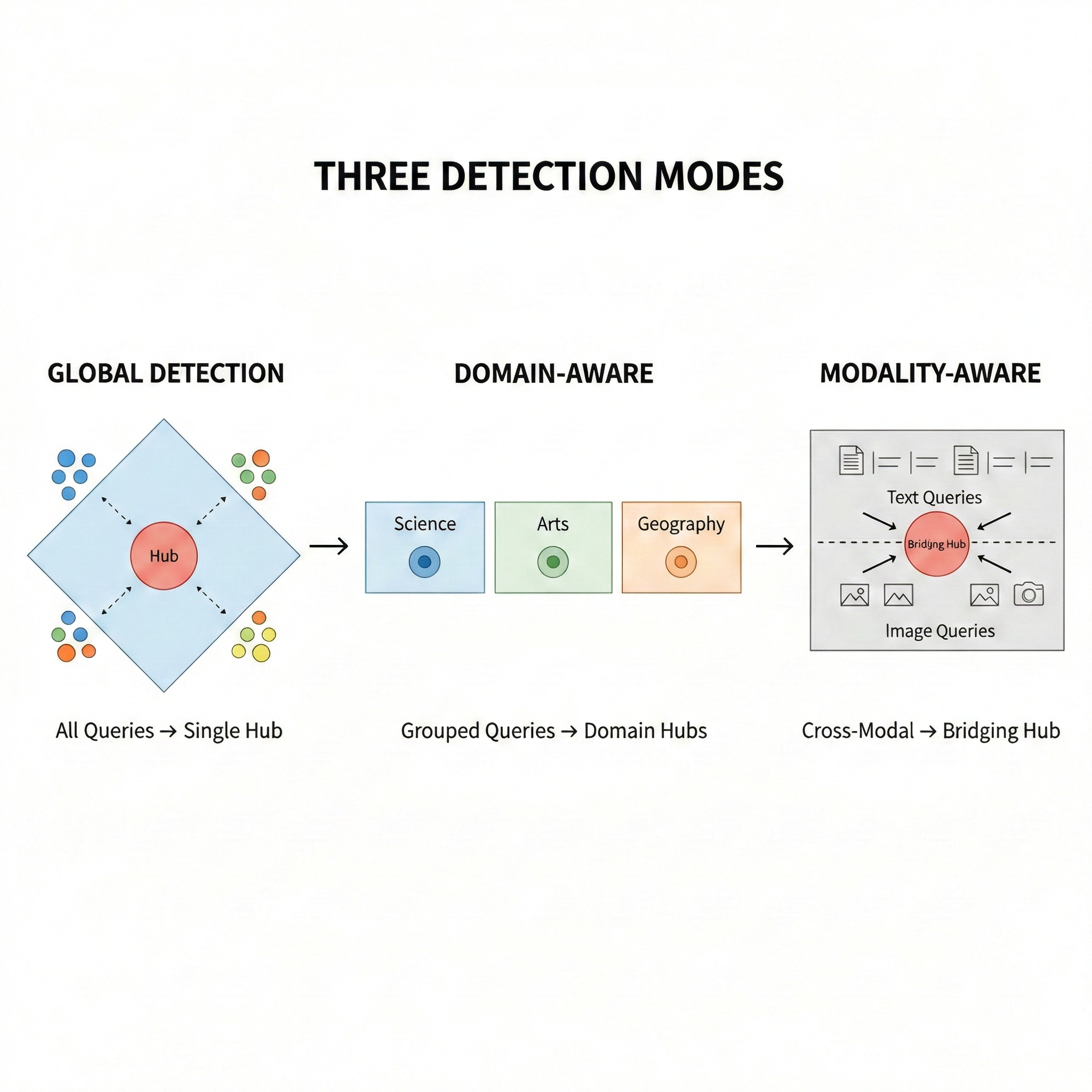}
\caption{Three detection modes: Global detection analyzes all queries together, Domain-Aware detection groups queries by semantic domains, and Modality-Aware detection handles cross-modal attacks.}
\label{fig:detection_modes}
\end{figure}

Standard global hubness detection may miss sophisticated attacks targeting specific semantic domains or exploiting modality boundaries. \hubscan provides specialized detection modes for these scenarios, as illustrated in Figure~\ref{fig:detection_modes}.

\subsection{Domain-Specific Hub Detection}

Domain-specific hub attacks aim to excel in retrieval within a certain semantic category while going undetected at a broader scale. Instead of establishing a universal hub for all inquiries, the attacker focuses influence on a singular notion (e.g., financial advice, medical advice, or legal interpretation), thereby circumventing detectors that depend on aggregate hubness statistics.

\textbf{Attack Characteristics}: Low global hubness avoiding detection by global threshold-based methods; extremely high hubness within a single semantic domain; disproportionate impact on targeted user segments; and applicability across modalities.

\textbf{Detection Framework}: We employ domain-aware hubness analysis where queries are categorized into semantic domains, and hubness is calculated independently for each domain. Domain assignment uses: (1)~metadata-based labeling when available, (2)~embedding-based clustering via MiniBatch K-Means, or (3)~hybrid assignment with clustering fallback.

For each domain, we calculate a topic-specific hubness score---the proportion of queries within that domain for which a retrieval item is included in the top-$k$ results. A retrieval item is flagged as suspicious when its hubness inside a single domain markedly exceeds its hubness across other domains (\textbf{contrastive detection}). We also compute a \textbf{concentration score} using the Gini coefficient based on the item's hubness distribution among domains:
\begin{equation}
G(d_i) = \frac{n + 1 - 2 \sum_{j=1}^{n} \frac{\sum_{k=1}^{j} p_{(k)}}{\sum_{k=1}^{n} p_{(k)}}}{n}
\end{equation}
where $p_{(k)}$ are the sorted normalized hub rates and $n$ is the number of domains. Values approaching 1 indicate that the majority of hubness is concentrated within a singular domain, typical of focused domain-specific attacks.

\textbf{Domain-Aware Scoring}: For each item, the scoring module produces: (1)~maximum domain-level hub z-score across all domains; (2)~dominant domain identifier; and (3)~hubness concentration score. This approach can identify globally benign but locally dominant items.

\subsection{Cross-Modal Hub Detection}

In multimodal retrieval systems (e.g., text--image or text--audio), adversarial hubs can exploit modality boundaries. A text-based retrieval item may be crafted to disproportionately appear in response to image-oriented queries, or vice versa. Such cross-modal dominance allows attackers to manipulate retrieval outcomes while evading detectors that operate within a single modality.

\textbf{Attack Characteristics}: The modality of the retrieval item differs from the dominant modality of the triggering queries; high retrieval frequency in cross-modal query settings; and evasion of single-modality hub detection mechanisms.

\textbf{Detection Framework}: If modality metadata is present for both queries and retrieved items, \hubscan tracks \emph{cross-modal hits}---when the query modality differs from that of the retrieved item. We calculate a \emph{cross-modal hub rate} for each item, normalized by the total number of searches, then compute a robust z-score over these rates to identify items receiving disproportionately many cross-modal hits. Items with high cross-modal z-scores receive an additive penalty term weighted in the final risk aggregate.

\begin{figure}[!t]
\centering
\includegraphics[width=1\linewidth]{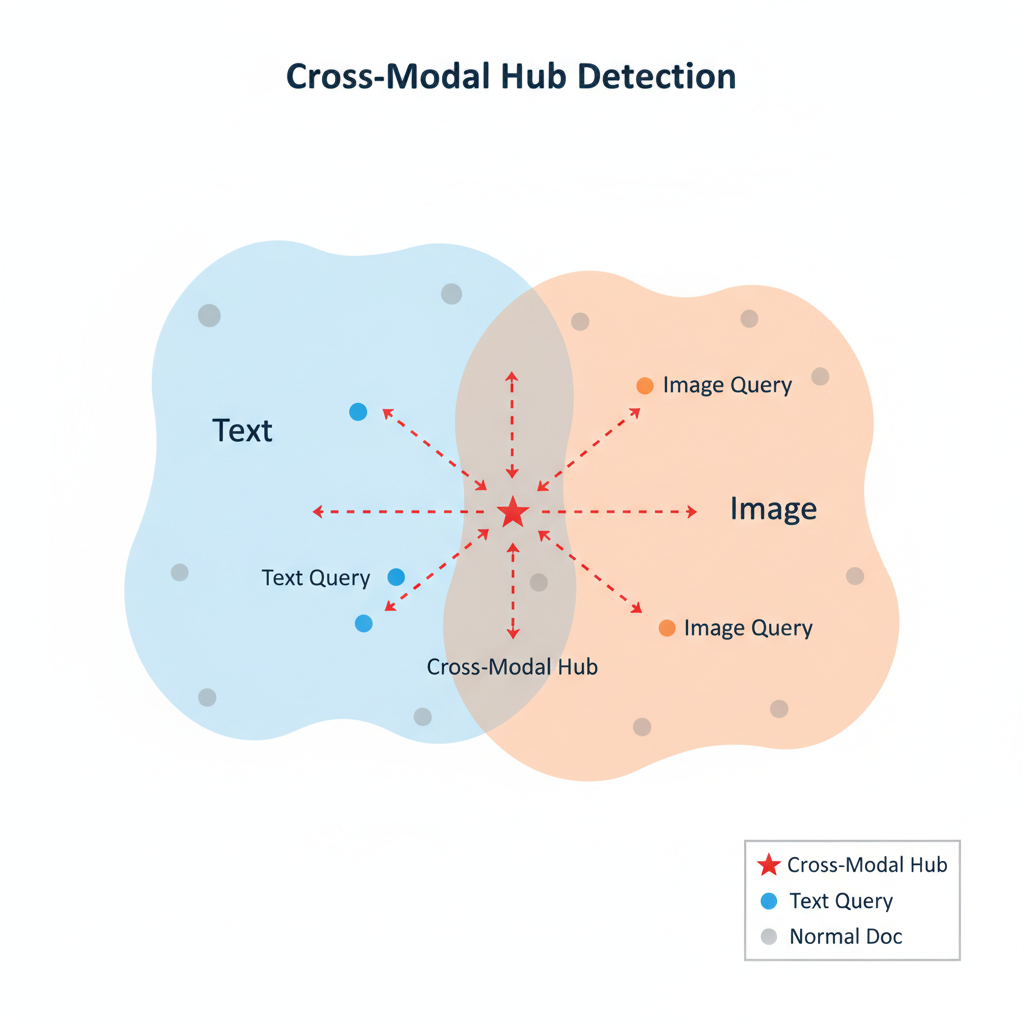}
\caption{Cross-modal hub detection: A hub (red star) positioned at the intersection of text and image modalities appears in top-$k$ results for queries from both modalities, exploiting the modality boundary.}
\label{fig:cross_modal_hub}
\end{figure}

Figure~\ref{fig:cross_modal_hub} illustrates how cross-modal hubs exploit modality boundaries.

\textbf{Modality-Aware Scoring}: For each item, the scoring module produces: (1)~cross-modal retrieval ratio; (2)~dominant query modality; and (3)~modality-adjusted hub score. The final score integrates global hubness with cross-modal behavior to identify items exploiting modality boundaries even when they appear benign within any single modality.

\section{Retrieval and Reranking Integration}
\label{sec:retrieval}

Effective hubness detection must operate within the retrieval and ranking pipelines used in production systems. \hubscan seamlessly connects with popular retrieval paradigms and reranking methodologies, aligning detections with user observations.

\subsection{Supported Retrieval Methods}

\textbf{Vector Similarity Search}: The standard retrieval method employs dense embeddings and approximate nearest neighbor search to obtain the top-$k$ most similar items. Similarity is calculated using cosine similarity or inner product. \hubscan integrates with widely used vector databases including FAISS, Pinecone, Qdrant, and Weaviate.

\textbf{Hybrid Search}: Hybrid retrieval integrates semantic (dense) similarity with lexical (sparse) matching, enabling systems to reconcile conceptual relevance with keyword overlap. This is particularly relevant for hubness detection because adversarial items optimized for semantic similarity may not dominate lexical search, and vice versa---enabling detectors to analyze whether an item behaves as a hub in one signal or across both.

\textbf{Lexical Search}: Lexical-only retrieval depends solely on keyword-centric scoring methods like BM25 or TF-IDF. This mode is beneficial for systems that emphasize keyword search or for isolating purely lexical manipulation techniques like keyword stuffing. In lexical-only pipelines, detectors reliant on embeddings (cluster spread, stability) are automatically deactivated.

\subsection{Reranking Support}

Many production systems apply a second-stage reranking model to refine initial retrieval results. Reranking typically operates on a larger candidate set returned by the retriever and produces a final, user-facing ranking. \hubscan facilitates customizable reranking workflows and assesses hubness according to post-reranking results, demonstrating that detection corresponds with the items finally presented to users.

\section{Evaluation}
\label{sec:evaluation}

\begin{figure*}[t!]
\centering
\includegraphics[width=1\linewidth]{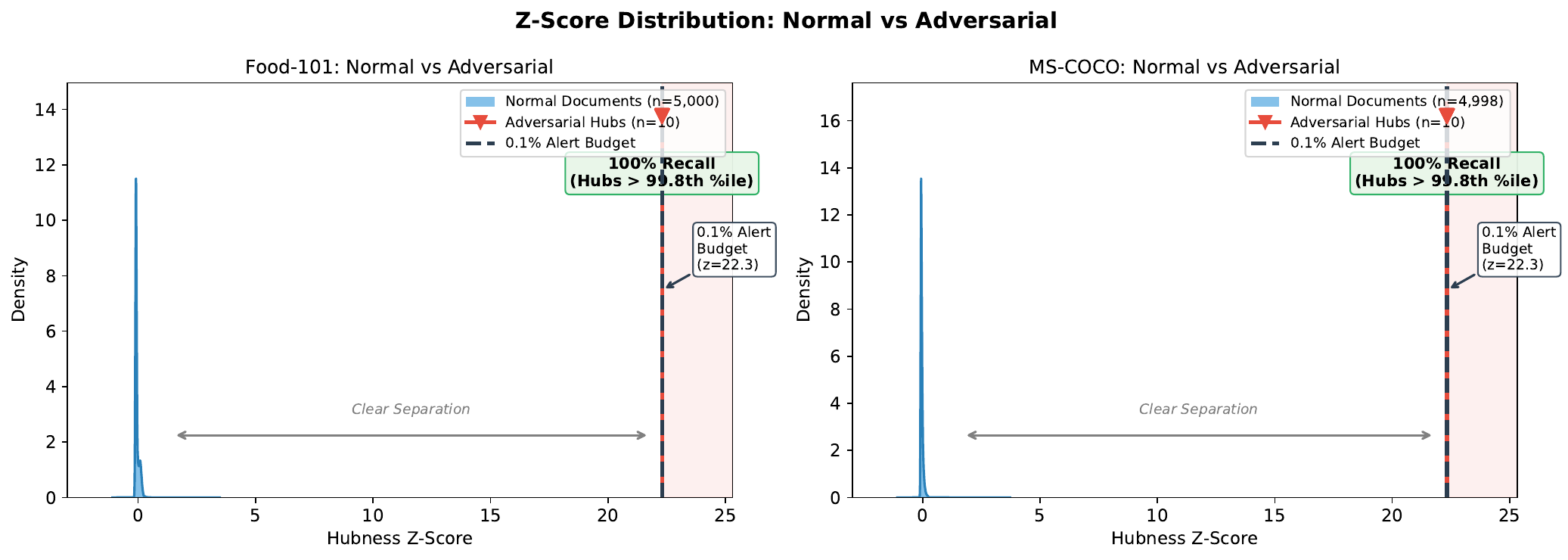}
\caption{Hubness score distribution for normal documents vs. planted adversarial hubs on Food-101 and MS-COCO. Normal items cluster near zero while adversarial hubs are extreme outliers (z-score $>$20). At a 0.1\% alert budget, all planted hubs rank above the 99.8th percentile, achieving 100\% recall with minimal false positives.}
\label{fig:zscore_distribution}
\end{figure*}

We evaluate \hubscan under the threat model outlined in Section~\ref{sec:methodology}, quantifying its efficacy in identifying adversarial \emph{hubs}---entities engineered to monopolize nearest-neighbor retrieval for numerous legitimate inquiries.

\subsection{Benchmark Construction}

Following standard adversarial ML evaluation methodology, we implemented the attack technique from Zhang~et~al.~\cite{zhang2024adversarial} to generate adversarial hubs, then evaluated \hubscan's ability to detect them. \textbf{Critically}, the attack and detection algorithms are independent: hub generation uses only embedding-space gradient optimization with no knowledge of \hubscan's statistical detection methods. This reflects a realistic threat model where attackers optimize embeddings without access to the defender's detection logic.

The Zhang methodology optimizes a hub embedding $h$ to maximize similarity to a target query set $\mathcal{T}$:
\begin{equation}
h^* = \argmax_{\|h\|=1} \sum_{q \in \mathcal{T}} \text{sim}(h, q)
\label{eq:zhang_objective}
\end{equation}
using momentum-based gradient ascent ($\mu{=}0.9$, $\eta{=}0.12$) with cosine temperature annealing. We evaluate two attack variants: \textbf{Universal} (random diverse queries; hub generalizes broadly) and \textbf{Domain-targeted} (domain-specific queries with repulsion term $\lambda_{\text{neg}}{=}3.0$ penalizing out-of-domain similarity).

To validate supplementary detection capabilities, we also utilize \textbf{centroid-based hubs} constructed as weighted averages of diverse document embeddings. Unlike gradient-optimized hubs, centroid hubs have artificial geometric placements and are sensitive to slight perturbations (denoted as \emph{brittle} hubs).

\subsection{Experimental Configuration}

Table~\ref{tab:datasets} summarizes the benchmarks. Image datasets use CLIP ViT-B/32 embeddings (512 dimensions), while FiQA uses Qwen3-0.6B embeddings (1024 dimensions). Universal hubs are optimized over 200 target queries for 1{,}000 gradient steps.

\begin{table}[h!]
\centering
\caption{Evaluation Benchmark Statistics}
\label{tab:datasets}
\small
\begin{tabular}{lccccc}
\toprule
\textbf{Benchmark} & \textbf{Docs} & \textbf{Queries} & \textbf{Hubs} & \textbf{Dim} & \textbf{Domains} \\
\midrule
Food-101~\cite{bossard2014food} & 5{,}010 & 25{,}000 & 10 & 512 & 101 \\
MS-COCO ~\cite{lin2014microsoft}  & 5{,}008 & 25{,}000 & 10 & 512 & 40 \\
FiQA ~\cite{maia201818}    & 5{,}005 & 648      & 5  & 1024 & --- \\
\bottomrule
\end{tabular}
\end{table}

Scanner configuration uses $k{=}20$ nearest neighbors, sampling 10{,}000 queries per scan via a mixed strategy (50\% random documents, 50\% cluster centroids).

\subsection{Evaluation Protocol}

We assess within a predetermined \textbf{alert budget}: with $N$ gallery items and a budget fraction $b$, we designate only the top-$K{=}\lceil b \cdot N \rceil$ highest-scoring items for manual examination. For example, $b{=}0.2\%$ on a $\sim$5{,}000-item gallery corresponds to reviewing the top 10 flagged items. We report \emph{recall} (fraction of adversarial hubs in top-$K$) and \emph{precision} (fraction of top-$K$ that are true hubs).

\subsection{Detection Results}

Table~\ref{tab:results} reports detection performance across datasets. Results are consistent, demonstrating that Zhang-style optimization yields similar statistical signatures regardless of the underlying domain or modality.

\begin{table}[h!]
\centering
\caption{Detection Performance (Full Configuration)}
\label{tab:results}
\small
\begin{tabular}{llcccc}
\toprule
\textbf{Dataset} & \textbf{Attack} & \textbf{Budget} & \textbf{K} & \textbf{Prec.} & \textbf{Recall} \\
\midrule
\multirow{2}{*}{Food-101}
  & Universal       & 0.2\% & 10 & 90\%  & 90\% \\
  & Universal       & 0.4\% & 20 & 50\%  & 100\% \\
\midrule
\multirow{2}{*}{MS-COCO}
  & Universal       & 0.2\% & 10 & 90\%  & 90\% \\
  & Universal       & 0.4\% & 20 & 50\%  & 100\% \\
\midrule
FiQA
  & Universal       & 0.1\% & 5  & 100\% & 100\% \\
\bottomrule
\end{tabular}
\end{table}

\textbf{Key findings}: With a 0.2\% alert budget ($K{=}10$), \hubscan accomplishes \textbf{90\% recall}, successfully identifying 9 out of 10 planted hubs. Expanding to $K{=}2H$ achieves \textbf{100\% recall} across tested datasets.

\subsection{Ablation Study: Detector Contributions}
\label{subsec:ablation}

We performed ablation tests to assess the contribution of each detector to universal hub detection. On Food-101, the Hubness-only configuration achieves 40\% recall at 0.1\% budget and 80\% at 0.2\%. Adding the \textbf{Cluster Spread detector} produces a \textbf{10--20 percentage point increase}: 50\% at 0.1\% and 100\% at 0.2\%.

This enhancement arises because universal hubs, optimized to attract diverse queries, demonstrate significant cluster dispersal. Universal hubs achieve a cluster spread of 0.92, while domain-targeted hubs obtain only 0.06---a \textbf{14$\times$ difference}. This explains why cluster spread is effective for universal attacks but not domain-targeted ones, which require domain-scoped scanning.

\subsection{Stability Detector: Catching Brittle Hubs}
\label{subsec:stability}

To validate the \textbf{stability detector's} coverage, we assess centroid-based hubs. Unlike Zhang's gradient optimization, centroid hubs are formed as weighted averages of document embeddings, making them naturally \emph{fragile}.

Stability is evaluated by introducing Gaussian noise ($\sigma{=}0.15$) to query embeddings. Zhang hubs show minimal hit change ($<$4\%), while brittle hubs show up to 68\% hit change \textbf{20$\times$ more unstable}. This verifies the stability detector's role: it detects centroid-based attacks that avoid the cluster spread detector, while Zhang-style hubs are detected by hubness and cluster spread signals. The modular detector architecture provides \emph{defense in depth} against multiple attack techniques.

\subsection{Domain-Scoped Scanning}
\label{subsec:domain_scoped}

Domain-targeted attacks present a distinct challenge: under global scanning with tight alert budgets, they can be pushed out by items with higher global hubness signals. We establish a benchmark where 15 benign universal hubs compete against 10 adversarial domain-targeted hubs.

During global scanning at $K{=}H$, \textbf{recall drops to 0\%}, i.e. all top-$K$ positions are held by the benign universal hubs. However, these domain hubs are not invisible: they achieve an AUC-ROC of 0.995. The issue is \emph{budget saturation}, not detection failure. \hubscan addresses this through \textbf{domain-scoped scanning}: filtering the query set to a specific domain before computing hubness statistics. When scoped to the target domain, during testing, domain-targeted hubs achieve \textbf{100\% recall}. This demonstrates the importance of layered defense: global scanning catches universal attacks, while domain-scoped investigation catches targeted attacks.

\subsection{Cross-Domain Generalization and Scaling Limits}

The results confirm that adversarial hubness is \textbf{modality-agnostic}: FiQA achieves \textbf{AUC-ROC of 1.0}, results in perfect separation with \textbf{100\% recall and 100\% precision at 0.1\% alert budget}.

To characterize the fundamental limits of statistical hub detection, we evaluate across varying adversarial corpus fractions (0.1\% to 30\%) on five domains: Food-101, MS-COCO (CLIP), FiQA (Qwen3), Code (CodeSearchNet with CodeBERT), Medical (PubMed abstracts with PubMedBERT). Table~\ref{tab:scalability} reports results.

\begin{table}[h!]
\centering
\caption{AUC-ROC by Adversarial Corpus Fraction (Hubness Z-Score Detector)}
\label{tab:scalability}
\small
\begin{tabular}{lccccc}
\toprule
\textbf{Domain} & \textbf{$\leq$2\%} & \textbf{5\%} & \textbf{10\%} & \textbf{20\%} & \textbf{30\%} \\
\midrule
Food-101  & 1.00 & 1.00 & 1.00 & 0.99 & 0.96 \\
MS-COCO   & 1.00 & 1.00 & 1.00 & 1.00 & 1.00 \\
FiQA      & 1.00 & 0.99 & 0.95 & 0.87 & 0.80 \\
CodeSearchNet      & 1.00 & 0.92 & 0.76 & 0.67 & 0.62 \\
PubMed   & 1.00 & 0.96 & 0.87 & 0.75 & 0.69 \\
\midrule
\textbf{Average} & \textbf{1.00} & \textbf{0.97} & \textbf{0.92} & \textbf{0.86} & \textbf{0.82} \\
\bottomrule
\end{tabular}
\end{table}

\textbf{Key finding}: Detection is perfect (AUC=1.0) when adversarial content comprises $\leq$2\% of the corpus, and remains highly effective ($>$0.9 AUC) up to 5\%. Detection degrades at higher fractions as adversarial items shift the distribution baseline. However, attacks of 10--30\% corpus fraction are unlikely threat scenarios that can be detected using simpler measures (corpus monitoring, ingestion rate alarms).

\subsection{Detection Signal Analysis}

Figure~\ref{fig:zscore_distribution} shows the hubness score distribution for normal items versus planted adversarial hubs. Normal items aggregate around zero, but adversarial hubs acquire z-scores of more than 20 standard deviations, making them significant outliers by any measure. During testing, at a 0.1\% alert budget (99.9\textsuperscript{th} percentile threshold), all planted hubs exceed the 99.8\textsuperscript{th} percentile, achieving \textbf{100\% detection with few false positives}. This percentile-based technique adapts across datasets and does not require manual threshold setting.

\section{Production Validation: Natural Hub Audit}
\label{sec:production_audit}

A fundamental concern for production deployment is establishing a baseline for the system's noise level: \textbf{what is the false positive rate of \hubscan in a clean environment with no active adversary?} To address this at a realistic web scale, we performed a natural hub audit on the MS MARCO~\cite{nguyen2016ms} passage retrieval corpus, an industry-standard dataset including 8.8 million documents sourced from actual Bing search results. We assessed a representative subset of \textbf{1 million passages}, proving validation at a production scale. We executed \hubscan using the same configuration employed for adversarial detection (mixed query sampling, $k{=}20$, complete multi-detector ensemble) and evaluated detection scores at 0.1\% and 0.2\% alert budgets to delineate operational workload.

\textbf{Score Separation.} Table~\ref{tab:clean_vs_adversarial_scores} demonstrates clear separation between clean and adversarial distributions. The 99\textsuperscript{th} percentile of clean data scored 2.3, while planted adversarial hubs from our evaluation (Section~\ref{sec:evaluation}) scored 13-17, resulting in a separation factor of \textbf{5.8$\times$}. The 99.9\textsuperscript{th} percentile of clean data (5.0) remains \textbf{2.7$\times$ lower} than adversarial hubs. This separation establishes distinct operational thresholds: score $>$10 demands investigation (adversarial range); score $<$5 indicates the clean baseline (99.9\textsuperscript{th} percentile).

\begin{table}[h!]
\centering
\caption{Score Distribution: Clean vs. Adversarial (1M Scale)}
\label{tab:clean_vs_adversarial_scores}
\small
\begin{tabular}{lccc}
\toprule
\textbf{Metric} & \textbf{Clean MS MARCO} & \textbf{Adversarial} & \textbf{Separation} \\
\midrule
99\textsuperscript{th} pct. & 2.3 & 13--17 & \textbf{5.8$\times$} \\
99.9\textsuperscript{th} pct. & 5.0 & 13--17 & 2.7$\times$ \\
\bottomrule
\end{tabular}
\end{table}

\textbf{Operational Workload.} At a 0.1\% alert budget, operators assess the top 1,000 highest-scoring documents per 1 million corpus. All highlighted passages received scores below 10.0, remaining beneath the adversarial range of 13--17. The 99\textsuperscript{th} percentile threshold (2.3) provides \textbf{5.8$\times$ separation} from adversarial content, while the 99.9\textsuperscript{th} percentile (5.0) maintains \textbf{2.7$\times$ separation}. Increasing the budget to 0.2\% (2,000 documents) preserved the same separation, with all scores remaining within the clean range. These represent expected false positives, documents scoring high relative to the clean baseline but far below true adversarial thresholds. The distinct score separation enables operators to implement an intuitive threshold (e.g., score $>$10) to quickly distinguish adversarial content from natural false positives, thus decreasing review time while preserving high detection sensitivity. This web-scale validation on 1 million real documents confirms operational feasibility for production RAG systems.

\section{Conclusion}
\label{sec:conclusion}

We have presented an in-depth study of hubness in multi-modal retrieval and a practical defense strategy. Hubs represent a potent attack vector: by exploiting the geometric properties of high-dimensional embeddings, a single malicious item can hijack retrieval results for a vast number of queries simultaneously \cite{zhang2024adversarial}. \hubscan is a comprehensive detection system addressing a critical security gap in the RAG ecosystem, where vector databases are increasingly targeted by sophisticated poisoning attacks.

Our multi-detector architecture combines statistical hubness detection, cluster spread analysis, stability testing, and domain/modality-aware detection. Evaluation demonstrates 90\% recall at 0.2\% alert budget and 100\% recall at $K{=}2H$, with adversarial hubs ranking above the 99.8\textsuperscript{th} percentile. In realistic threat models where adversaries introduce a limited number of high-impact hubs ($\leq$2\% of the corpus), the hubness z-score detector alone achieves perfect detection.

\subsection{Limitations}

Current limitations include:
\begin{itemize}
    \item \textbf{Query Distribution Dependence}: Detection accuracy relies on representative query sampling or queries from production.
    \item \textbf{Adaptive Adversaries}: Attackers aware of detection mechanisms may attempt evasion.
    \item \textbf{Low-Volume Attacks}: Attacks with minimal hub rates may evade statistical detection.
    \item \textbf{Statistical Limits}: The hubness z-score detector's effectiveness degrades when adversarial content exceeds 10\% of the corpus, requiring corpus monitoring for large-scale attacks.
\end{itemize}

\subsection{Future Directions}
To improve \hubscan, future work will include:
\begin{itemize}
    \item Real-Time Detection: Adjusting \hubscan to flag potential hubs during the indexing process, by pre-calculating hubness score against set of queries.
    \item Intent Scanning: Extending \hubscan with an intent scanner to search for malicious content inside flagged items (e.g. prompt injections or hidden content).
    \item Adaptive Adversaries: Explore how \hubscan performs against attackers that adapt their planted items to bypass hubness-based detection, a potential strategy can be planting items that will disrupt such attempts.
\end{itemize}

\subsection{Open Source Release}
\label{sec:availability}

We release \hubscan as open-source software to enable security auditing of production RAG systems. The release includes the complete implementation, adapters for popular vector databases, all detection algorithms described in Section~\ref{sec:detection}, and reproduction scripts for our benchmarks.

\noindent\textbf{Repository:} \url{https://github.com/cisco-ai-defense/adversarial-hubness-detector}

\appendix
\section{Ethical Considerations}
\label{sec:ethics}


This work presents \hubscan, a defensive tool designed to detect adversarial hubness attacks in RAG systems. We address the following ethical considerations:

\textbf{Dual-Use Concerns.} While we use hub creation techniques from previous works to contextualize our defense, we do not introduce any new attack strategies. Our primary focus is on detection and mitigation. The attack strategies we assess have already been publically described in peer-reviewed literature.

\textbf{Responsible Disclosure.} The real-world vulnerabilities discussed (Microsoft Copilot, GeminiJack) were discovered and disclosed by independent security researchers prior to our work. We cite these incidents to motivate the need for hubness detection, not to enable new attacks.

\textbf{Benchmark Construction.} Our evaluation benchmarks use publicly available datasets (Food-101, MS-COCO, FiQA) with synthetically planted adversarial hubs. No real user data or production systems were involved in our experiments.

\textbf{Intended Use.} \hubscan is designed for security practitioners to audit and protect RAG deployments. We encourage responsible use for defensive purposes and discourage any application that could harm users or systems.

\section{Open Science}
\label{sec:open_science}

We are committed to open and reproducible research:

\textbf{Open-Source Software.} \hubscan is released under the Apache 2.0 license at https://github.com/cisco-ai-defense/adversarial-hubness-detector.
The repository includes the complete detection framework, all detector implementations, and adapter interfaces for major vector databases (FAISS, Pinecone, Qdrant, Weaviate).

\textbf{Benchmark Datasets.} Our evaluation benchmarks, including the adversarial hub generation scripts following Zhang et al.~\cite{zhang2024adversarial}, are included in the repository under \texttt{benchmarks/}. This enables researchers to reproduce our results and evaluate new detection methods.

\textbf{Evaluation Scripts.} All scripts used to generate the results in this paper are provided, including hub generation, detection execution, and metric computation. Configuration files for each experiment are included.

\textbf{Documentation.} Comprehensive documentation covers installation, configuration, API usage, and extension development. Example notebooks demonstrate common use cases.

\textbf{Reproducibility.} Random seeds are fixed for all stochastic components. Hardware requirements and expected runtimes are documented. We provide pre-computed embeddings for the benchmark datasets to reduce computational barriers to reproduction.

{\footnotesize \bibliographystyle{acm}
\bibliography{references}}

\end{document}